 \documentstyle[prl,aps]{revtex}  
\begin{document}
\draft


 \twocolumn[\hsize\textwidth\columnwidth\hsize  
 \csname @twocolumnfalse\endcsname              

\title{Single-Particle and Collective Motion   
for Proton-Rich Nuclei \\in the Upper $pf$ Shell}
\author{Yang Sun$^{(1,2)}$, Jing-ye Zhang$^{(1)}$,
Mike Guidry$^{(1)}$, 
Jie Meng$^{(3)}$ and Soojae Im$^{(3)}$ 
}
\address{
$^{(1)}$Department of Physics and Astronomy, University of Tennessee,
Knoxville, Tennessee 37996 \\
$^{(2)}$Department of Physics, Xuzhou Normal University,
Xuzhou, Jiangsu 221009, P.R. China\\
$^{(3)}$Nuclear Theory Laboratory, Department of Technical Physics, 
Peking University, Beijing 100871, P.R. China}

\date{\today}
\maketitle

\begin{abstract}
Based on available experimental data,
a new set of Nilsson parameters is proposed
for proton-rich nuclei with proton or neutron numbers $28\leq N\leq 40$.
The resulting single-particle spectra are compared with those 
from relativistic and non-relativistic mean field
theories.
Collective excitations in some even--even proton-rich nuclei
in the upper $pf$ shell are investigated  
using the Projected Shell Model with the new Nilsson basis.
It is found that the regular bands are sharply disturbed by band crossings 
involving $1g_{9/2}$ neutrons and protons. 
Physical quantities for exploring the nature of the band disturbance
and the role of the $1g_{9/2}$ single-particle are predicted, which 
may be tested by new experiments with radioactive beams. 
\end{abstract}


 ]  

\narrowtext

\bigskip
The nuclear shell model has been successful in the description of nuclear
structure. Thanks to the increasing power of computation, exact   
diagonalizations in the full $pf$ shell 
has become possible
in recent years \cite{Ca94}.  
An immediate application has been seen in nuclear astrophysics 
\cite{Ca99,La00}, 
where knowledge about detailed nuclear structure is important 
in understanding the nuclear processes that govern those violent astrophysical
phenomena such as nova and supernova explosions, or X-ray bursts.  

Nuclear structure information is thought to be important also 
in the study of the nuclear processes occurring on the astrophysical
rapid proton capture or rp-process \cite{WW81,Sch98}, which 
may be relevant to nova explosions or X-ray bursts. 
The rp-process path lies close to the proton drip line in 
the chart of nuclides. 
There, compound nuclei are
formed at very low excitation energies and therefore at low level densities,
which does not justify Hauser-Feshbach calculations. 
Thus, detailed nuclear structure information is required
when studying the nuclear processes. 
One hopes that radioactive beams will provide us 
with the information eventually,
but one has to rely on theoretical calculations at present. 
To obtain detailed nuclear structure, advanced shell model 
diagonalization methods, which can give explicitly spectroscopy 
and matrix elements for
all kinds of nuclei (even-even, odd-A and odd-odd), are
of particular importance.

The Projected Shell Model (PSM) \cite{HS95}
is a shell model diagonalization carried out in a projected 
space determined by a deformed
Nilsson--BCS basis.
This kind of shell model truncation 
is highly efficient if the single-particle (SP) basis is realistic
because the basis already
contains many correlations \cite{HS95}.
It has been shown that the PSM can describe
the spectra and electromagnetic transitions in normally deformed \cite{HS95},
superdeformed \cite{SD190,SD60}, and
transitional nuclei \cite{trans}.
One can further calculate the nuclear matrix elements
for astrophysical processes such as
direct capture and decay rates.
One advantage of the PSM 
is that it can easily handle heavy, well-deformed nuclear systems.  
This can be important for the structure study of significantly deformed
nuclei ($Z=36-40$) on the rp-process path, for which the current large-scale 
shell model diagonalizations are not feasible. 

As an initial step for a PSM study, 
a reliable Nilsson model calculation \cite {nil,tod} 
is required to build the  
projected space. 
The Nilsson 
model has been used widely in nuclear structure studies.
Its ``standard'' set of parameters
${\kappa}$ and ${\mu}$ \cite {tod} has been
quite successful in describing the SP structure for stable
nuclei across the whole chart of nuclides.
However, previous work has shown that
the standard Nilsson SP energies 
are no longer realistic for the neutron-rich region
\cite{Sn132}.    
Thus, an adjustment of these parameters is necessary for unstable nuclei.

Experimental SP states  
above the $N=28$ shell closure can be read from the 
low-lying states
of $^{57}$Ni \cite{Bh92} and $^{57}$Cu \cite{Zh96}, 
the nearest neighbors of the doubly magic nucleus
$^{56}$Ni. 
These low-lying $2p_{3/2}$, $1f_{5/2}$ and $2p_{1/2}$ states are 
well characterized as pure SP levels \cite{Re98}.  
According to Ref. \cite{Ru99}, the neutron $1g_{9/2}$ state lies 3.7 MeV
above the $2p_{3/2}$ orbital. For the proton $1g_{9/2}$ state, 
the best experimental information available is the observation 
of a low-lying ${9\over 2}^+$ level 
in $^{59}$Cu \cite{iso} and 
$^{61}$Cu \cite{Vi99}. We may reasonably assume this level to be the
$1g_{9/2}$ state because of its unique parity.    
Taking the deformation of $^{59,61}$Cu \cite{mol} into account, 
the position of the proton SP 
$1g_{9/2}$ state can be estimated to lie 3.15 MeV
above the $2p_{3/2}$ orbital.

When compared with these data, 
the standard Nilsson parameterization \cite{tod}
produces energies for the $1f_{5/2}$, $2p_{1/2}$ and $1g_{9/2}$  
orbits that are too high relative to the $2p_{3/2}$ orbit. 
To reproduce experimental data, 
one must reduce
the strength of the spin--orbit interaction $\kappa$ for $N=3$
substantially from the standard value 
because of the observed
smaller separation between $2p_{1/2}$
and $2p_{3/2}$ levels, and because of the position of $1f_{5/2}$ orbital.
On the other hand,
the pair of $g$-orbitals require a larger value of $\kappa$
for $N=4$ to position the $1g_{9/2}$ orbital properly. 
In Table I, we summarize the adjusted proton and neutron
$\kappa$ and $\mu$ parameters for the $N$ = 3 and 4 shells that
best reproduce the available
data.
The standard values for the $N$ = 2 
shell are also displayed without modification.

Relativistic Mean Field (RMF) theory \cite{Ring96} with
nonlinear self-interactions between mesons has
been used in many studies
of low-energy phenomena in
nuclear structure.
It has been extended recently to allow coupling between bound states
and the continuum by the pairing force \cite{Meng98}. 
In the RMF theory, the spin-orbit interaction
arises naturally 
as a result of the Dirac structure of the nucleons.
As discussed above, the spin--orbit interaction is one of the important factors
to give a correct SP energies. 
Thus, it is relevant to consider the relation of the SP
Nilsson spectrum to that of the RMF.
Here, 
two typical interactions, NL1 and NL3, are used. 
The latter interaction is suitable also for nuclei
away from the $\beta$-stability valley \cite{La98}. 
For comparison, SP energies from the non-relativistic 
Hartree-Fock calculations with the Skyrme interactions (SHF) 
(see Ref. \cite{SHF} and references therein) 
are also presented.

In Fig. 1, theoretical SP states (from the new and the standard \cite {nil,tod} 
Nilsson parameterizations, from SHF with SkM$^*$ and SIII forces,
and from RMF with NL1 and NL3 interactions) are compared with data.
It is obvious that the standard Nilsson parameters produce a large
spin-orbital splitting for $2p_{1/2}$ and $2p_{3/2}$,
and high excitation energy for $1f_{5/2}$ 
while the new set of parameters
reduces these values.
The SHF and RMF calculate the SP states for $^{56}$Ni.  
Without specific parameter adjustment for this mass region,  
SP levels of the RMF with NL3 for the $p$ and $f$ orbitals
are found to be reasonably close to the data.
However, the SHF with the SkM$^*$ force produces rather wrong
positions for all levels considered.  
In all the SHF and RMF calculations, the positions of both neutron and proton 
$1g_{9/2}$ orbital are much too high. 

The new Nilsson parameters should represent a
better basis from which more sophisticated wave functions can be
constructed. Therefore, we may test the new parameters
by employing them in calculations that have a direct
connection with measured collective spectra.
For such a test, we shall employ the PSM
to calculate the yrast bands of some even--even nuclei
for which limited data are available for comparison:
$^{62,66}$Zn and $^{64,66}$Ge. These nuclei lie on the rp-process path, 
and $^{64}$Ge
is a waiting point nucleus \cite{Sch98} that was used as a test case 
for the recently proposed 
Quantum Monte Carlo Diagonalization Method (QMCD) \cite{QMCD}. 
As we shall see, the new Nilsson SP states discussed above 
can modify substantially the level spacings, position
and sharpness of band crossing, and electromagnetic transition properties
along a level sequence. 

In PSM calculations of the collective states relevant here,
the projected multi-quasiparticle (qp) 
space consists of 0-, 2-qp and 4-qp 
states for even--even
nuclei, typically with a dimension of 50.
This small shell model basis is sufficiently rich 
that the quality of the calculations is comparable to 
those from large-scale shell model diagonalizations \cite{Ha99}.  
For the SP basis we use three full major shells:
$N$ = 2, 3, and 4 for both neutrons and protons. 
This is the same size SP basis as employed in Ref. \cite{SD60}. 
The deformation parameters are taken from Ref. \cite{mol} as follows:
$\varepsilon_2=0.167$ and $\varepsilon_4=-0.020$ for $^{60}$Zn, 
$\varepsilon_2=0.192$ and $\varepsilon_4= 0.013$ for $^{62}$Zn, 
$\varepsilon_2=0.200$ and $\varepsilon_4= 0.047$ for $^{64}$Ge, 
and $\varepsilon_2=0.208$ and $\varepsilon_4= 0.067$ for $^{66}$Ge. 
The Hamiltonian is the usual quadrupole-quadrupole plus monopole pairing
form, with quadrupole pairing included \cite{HS95}. 
The strength of the 
quadrupole-quadrupole force in the Hamiltonian is determined 
self-consistently, and the monopole pairing
strength is the same as that in Ref. \cite{SD60}. 
For the four nuclei calculated in this paper, 
the ratio of quadrupole pairing to monopole pairing strength
is fixed at 0.30. 

In Fig.\ 2, yrast band (lowest energy state 
for given spin) 
transitional energies $E(I)-E(I-2)$ are plotted as a function of spin $I$.
It is obvious that the calculations employing
the new set of Nilsson parameters
reproduce the data very well, while those with the standard set
of Nilsson parameters determined in the stability valley are in poorer
agreement.
In all the four nuclei, the standard set of Nilsson parameters
gives too large level spacings for low-spin states, 
leading to the excitation energies that are too high.
Note that the two calculations are performed with the same conditions
except for different SP bases.  
It is the change in SP states that gives rise
to the different results for the yrast spectrum.

Following the yrast band in a nucleus in Fig. 2, one observes a sudden drop 
in $E(I)-E(I-2)$ 
at spin $I=8$ or 10, 
which corresponds to a backbending
in the moment of inertia for the system \cite{HS95}.
For these $N\sim Z$ nuclei, neutron and proton Fermi levels are 
surrounded by orbits with the same
Nilsson quantum numbers. Therefore, bands built on the neutron and
proton $1g_{9/2}$ intruder orbits can have a similar probability to be
the first that crosses the ground band and becomes a major part of the
yrast band.    
From analysis of the wave functions, we find that the sudden drop
is caused by such band crossings. 
The crossing bands have either 2-neutron $1g_{9/2}$ or 2-proton $1g_{9/2}$
configurations.

Effects of the band crossing can be seen more clearly 
by looking at the reduced transition rates B(E2).
In the B(E2) calculations, 
the effective charges used in this paper 
are 0.5e for neutrons and 1.5e for protons, 
which are the same as those used in previous work and in other
shell models \cite{Ha99}, and similar to those in the QMCD. 
For a comparison, 
our results for the first two transitions in 
$^{64}$Ge (see Fig. 3a) are 
close to those obtained in 
the QMCD \cite{QMCD} (B(E2; $2^+ \rightarrow 0^+) = 0.050 (e^2b^2)$ 
and B(E2; $4^+ \rightarrow 2^+) = 0.065 (e^2b^2)$).
We emphasize that employment of different effective charges 
can modify the absolute values, but the essential spin dependence 
is determined by the wave functions. 
In Fig. 3a, the sudden drop in
the B(E2) values at spin $I=8-10$ 
is consequence of band crossings discussed above.
The drop indicates that the transition rate is sharply reduced 
by the structure change in the
yrast band wave function.
In \cite{QMCD}, states in $^{64}$Ge were calculated up to $I=4$.
It would be an important comparison if the QMCD results could be extended 
beyond that spin. 
It is interesting to point out that 
our prediction for the crossings occurs exactly 
at the excited states where energy spectra measured to date have terminated
(see Fig. 2).
Thus the band crossing predictions 
may explain why the experimental measurements have not seen 
higher spin states. 

The gyromagnetic factor (g-factor) is the quantity 
most sensitive to the SP components in wave functions 
as well as to their interplay with collective degrees of freedom.  
Because of the intrinsically opposite signs of the neutron and proton
$g_s$, 
a study of g-factors enables determination of the microscopic structure
for underlying states. For example, variation of g-factors 
often is a clear indicator for a 
SP component that strongly influences the total wave function.
In the calculations we use for $g_l$ the free values
and for $g_s$ the free values damped by the usual 0.75 factor. 
The results are presented in Fig. 3b.

Rather similar behavior is seen for all the four nuclei 
for spin states before band crossing:  
The g-factor values are close to the collective value of $Z/A$
and tend to increase slightly as a function of spin.  
However, the pattern diverges at band crossing.
The g-factors of the two $N=Z$ nuclei ($^{60}$Zn and $^{64}$Ge) 
jump suddenly to a value near 1, while those of the two 
$N=Z+2$ nuclei ($^{62}$Zn and $^{66}$Ge)  
drop to zero.
By analysis of the wave functions, we have found that this interesting
behavior is determined by whether the proton or neutron 
$1g_{9/2}$ 2-qp states is lower in energy after band crossing. 
For $N=Z$ nuclei, proton $1g_{9/2}$ 2-qp states dominate the wave functions
after band crossing, thus increasing the g-factors. 
For $N=Z+2$ nuclei, neutron $1g_{9/2}$ 2-qp states become lower due to
different neutron shell fillings, leading to very low values of the g-factors. 
 
Similar band crossing pictures should be common for this mass
region, and may also be seen in the neighboring odd-A
and odd-odd nuclei. 
Successful measurement of the B(E2) values and 
g-factors before and after band crossing 
would test our PSM predictions as well as the Nilsson SP
states proposed in this paper. 
We hope that recently developed techniques \cite{g-factor} 
in combination with
radioactive beams can permit such measurements. 

In summary, a new set of Nilsson parameters is proposed for proton-rich nuclei
beyond $^{56}$Ni.  
This new
set of parameters can be employed 
for proton-rich nuclei with proton and number numbers $N = 28 - 40$. 
Our results suggest that 
improved SP structures 
may be necessary for this mass region
in SHF and RMF theories. 
Available data for
yrast bands in
some even--even nuclei on the rp-process path are well reproduced by
PSM calculations with the new set of Nilsson states
as a basis.
Distinct signatures for the role of 
neutron and proton $1g_{9/2}$ orbitals are seen 
in the band crossings and the electromagnetic transitions along 
the yrast bands. 
We conclude that  
precise positions of the $1g_{9/2}$ SP orbitals are very important
for any realistic calculations.
Predictions of the present paper concerning the $1g_{9/2}$ band crossings, 
B(E2) values, and the remarkably different behaviors of the g-factors 
at the crossings await future experimental tests.

                            Research at the University of
                            Tennessee is supported by the U.~S. Department
                            of Energy through Contract No.\
                            DE--FG05--96ER40983.
This work was partially sponsored by the National Science Foundation
      of China under Project No. 19847002 and 19935030, and by SRF for ROCS,
      SEM, China.

\baselineskip = 14pt
\bibliographystyle{unsrt}

\begin{figure}
\caption{SP states for neutrons and protons.
Experiment and calculations from Nilsson
model with new parameter set, from the standard set,
from SHF with SkM$^*$ and SIII forces, 
and from RMF with NL1 and NL3 parameterizations.}
\label{figure.1}
\end{figure}

\begin{figure}
\caption{Transition energies $E(I)-E(I-2)$ along 
the yrast band in $^{60}$Zn, $^{62}$Zn, 
$^{64}$Ge and $^{66}$Ge from
experimental data \protect\cite{iso,Zn60} (filled triangles) and 
PSM calculations with the new set of Nilsson parameters (open
circles),
and with
the standard Nilsson parameters (open squares).
}
\label{figure.2}
\end{figure}

\begin{figure}
\caption{B(E2) (in unit of $e^2b^2$) and g-factors in 
$^{60}$Zn, $^{62}$Zn,
$^{64}$Ge and $^{66}$Ge from 
PSM calculations with the new set of Nilsson parameters.
}
\label{figure.3}
\end{figure}

\begin{table}[h]
\begin{center}
\caption{Nilsson parameters $\kappa$ and $\mu$ for the major shells $N$ = 2, 3
and 4.} 
\vspace{0.5cm}
\begin{tabular}{c|c|c|c|c|c|c|c|c}
$N$&$\kappa_{p}$& new $\kappa_{p}$&$\mu_{p}$&new $\mu_{p}$
&$\kappa_{n}$& new $\kappa_{n}$&
$\mu_{n}$&new $\mu_{n}$\\
\hline
2 &0.105&      &0.0  &      &0.105&      &0.0  &      \\
\hline
3 &0.090&0.039 &0.300&0.222 &0.090&0.035 &0.250&0.293 \\
\hline
4 &0.065&0.087 &0.570&      &0.070&0.097 &0.390&      \\
\end{tabular}
\end{center}
\end{table}

\end{document}